# Design and Performance Analysis of Unified Reconfigurable Data Integrity Unit for Mobile Terminals

L.Thulasimani
Department of ECE
PSG College of Technology,
Coimbatore-641004,India
.

M.Madheswaran
Centre for Advanced Research, Dept. of ECE
Muthayammal Engineering College
Rasipuram-637408, India
.

*Abstract*—**Security has become one of the major issue in mobile services. In the development of recent mobile devices like Software Defined Radio (SDR) secure method of software downloading is found necessary for reconfiguration. Hash functions are the important security primitives used for authentication and data integrity. In this paper, VLSI architecture for implementation of integrity unit in SDR is proposed. The proposed architecture is reconfigurable in the sense it operates in two different modes: SHA-192 and MD-5.Due to applied design technique the proposed architecture achieves multi-mode operation, which keeps the allocated area resource at minimized level. The proposed architecture also achieves high-speed performance with pipelined designed structure. Comparison with related hash function implementation have been done in terms of operating frequency, allocated-area and area-delay product. The proposed Integrity Unity can be integrated in security systems for implementation of network for wireless protocol, with special needs of integrity in data transmission.**

*Index Terms*— **SDR, reconfigurability, SHA-192, Unified architecture, Hardware utilization**

1. INTRODUCTION

Cryptographic hash functions have been wide applied in science of information security. It protects data from theft or alteration and it can also be used for user authentication. Modern cryptography concerns itself with confidentiality, integrity, non-repudiation, and authentication. There is current and growing interest in universal terminals (multi services, multi networks) for wireless networks. The technical approach to these universal terminals includes developing reconfigurable terminals. The reconfigurable terminals can change their hardware configuration and can support multi-operation modes. This idea of reconfigurablility leads to the development of software radio techniques which requires secure software downloading for reconfiguration.

Hardware architecture for high performance AES algorithm has been implemented for encryption process which is useful for SDR terminals[1].Also radio security module that offers a SDR security architecture that enables separate software and hardware certification is being developed[2].Security encipherment is achieved using the characteristics of the Field Programmable Gate Array, which allows the system to be arranged in a variety of different layouts[3].Cryptographic components are also exchanged for secure download. It includes the possibility to change any of the cryptographic components employed [4]-[5]. In this paper, reconfigurable hardware architecture has been proposed with an aim to provide secure download in SDR terminals. Also the area utilization of proposed architecture is analyzed with an aim to have optimized area and power consumption.

2. MD-5 AND SHA-1 ALGORITHM

*2.1. MD5 Algorithm*

MD5 [6] was introduced in 1992 by Professor Ronald Rivest. It calculates a 128-bit digest for an arbitrary l-bit message. It is an enhanced version of its predecessor MD4.The algorithm could be described in two stages: Pre-processing and hash computation. Preprocessing involves padding a message, parsing the padded message into m-bit blocks, and setting initialization values to be used in hash computation. The final hash value generated by the hash computation is used to determine the message digest.

*1. Append Padding Bits* The b-bit message is padded so that a single 1 bit is appended to the end of the message, and then 0 bits are appended until the length of the message becomes congruent to 448, modulo 512.

*2. Append Length* A 64-bit representation of b is appended to the result of the padding. The resulting message has a length that is an exact multiple of 512 bits. This message is denoted here as Y.

*3. Initialize MD* Buffer Let A, B, C, D be 32-bit registers. These registers are initialized to the following values in hexadecimal, low-order bytes first:Word A: 01234567 B: 89abcdef Word C: fedcba98 Word D: 765432

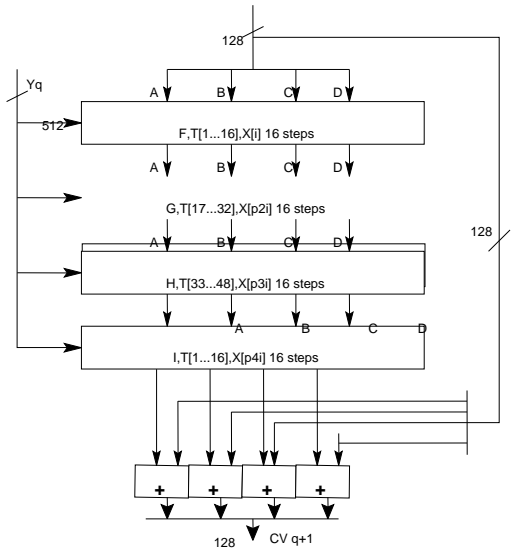

**Figure 1. Compression function HMD5**

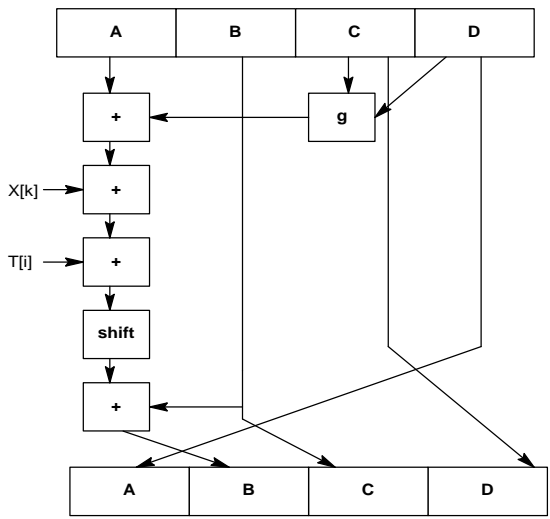

**Figure 2. Operation of single step of MD5**

*4. Process Message in 16-Word Blocks* This is the heart of the algorithm, which includes four rounds of processing. It is represented by HMD5 in Fig.1 and its logic is given in Fig 2. The four rounds have similar structure but each uses different auxiliary functions F, G, H and I.

$F(X,Y,Z) = (X$ and $Y)$ or $((notx)$ and $Y)$
$G(X,Y,Z) = (X$ and $Z)$ or $(Y$ and $(notZ))$
$H(X,Y,Z) = X$ xor $Y$ xor $z$
$I(X,Y,Z) = Y$ xor $(X$ or $(notZ))$

Each round consists of 16 steps and each step uses a 64-element table T [1 ... 64] constructed from the sine function. Let T[i] denote the i-th element of the table, which is equal to the integer part of 232 times abs(sin(i)), where i is in radians. Each round also takes as input the current 512-bit block Yq and the 128-bit chaining variable CVq. An array X of 32-bit words holds the current 512-bit Y. For the first round the words are used in their original order.

The following permutations of the words are defined for rounds 2 through 4:

$?2(i) = (1 + 5i)$ mod 16
$?3(i) = (5 + 39)$ mod 16
$?4(i) = 7i$ mod 16

The output of the fourth round is added to the input of the first round (CV,) to produce CVq+l.

*5. Output* After all L 512-bit blocks have been processed, the output from Lth stage is the 128-bit message digest. Fig 2 shows the operations involved in single step. The additions are modulo 232. Four different circular shift amounts S is used each round and are different from round to round. Each step is of the following form,

A -> D
B -> B + ( ( A + Funs ( B , C , D ) + X [ K l + T [ I ] ) < < s )
C -> B
D -> C

### 2.2. The SHA-1 Algorithm

The Secure Hash Algorithm was developed by National Institute of Standards and Technology (NIST) and published as a federal information processing standard in 1993[7]. It calculates a 160-bit digest for an arbitrary l-bit message. Preprocessing is done same as in MD5 except that an extra 32-bit register E is added with an initial value of C3D2E1F0. Other registers are assigned with higher order bytes first. For each block, it requires 4 rounds of 20 steps, resulting in a total of 80 steps, to generate the message digest. Fig.3 shows the SHA-1 compression function [8].

*Functions* A sequence of logical functions f0, f1,..., f79 is used in the SHA-1. Each ft, $0 <= t <= 79$, operates on three 32-bit words B, C, D and produces a 32-bit word as output. ft(B,C,D) is defined as follows, for words B, C, D,

$ft(B,C,D) = (B$ and $C)$ or $((not B)$ and $D)$, for $0 <= t <= 19$
$ft(B,C,D) = B$ xor $C$ xor $D$, for $20 <= t <= 39$
$ft(B,C,D) = (B$ and $C)$ or $(B$ and $D)$ or $(C$ and $D)$, for $40 <= t <= 59$
$ft(B,C,D) = B$ xor $C$ xor $D$, for $60 <= t <= 79$

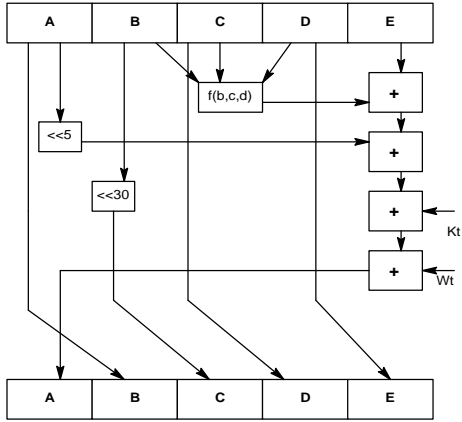

**Constants** A sequence of constant words K(0), K(1), ... , K(79) is used in the SHA-1. In hex these are given by

$Kt = 5A827999$ ( $0 <= t <= 19$)
$Kt = 6ED9EBA1$ ($20 <= t <= 39$)
$Kt = 8F1BBCDC$ ($40 <= t <= 59$)
$Kt = CA62C1D6$ ($60 <= t <= 79$)

*Computing the Message Digest*

The message digest is computed using the final padded message. The computation uses two buffers, each consisting of five 32-bit words, and a sequence of eighty 32-bit words. The words of the first 5-word buffer are labeled A, B, C, D, E. The words of the second 5-word buffer are labeled H0, H1, H2, H3, H4. The words of the 80-word sequence are labeled W0, W1... W79. A single word buffer TEMP is also employed. To generate the message digest, the 16-word blocks M1, M2... Mn is processed in order. The processing of each Mi involves 80 steps. Single step operation of SHA-1 is shown in Fig.4. Before processing any blocks, the {Hi} are initialized as follows in hex:
H0 = 67452301, H1 = EFCDAB89, H2 = 98BADCFE, H3 = 10325476, H4 = C3D2E1F0.
Now M1, M2... Mn is processed. To process Mi, the following procedure can be executed:

*a. Divide Mi into 16 words W0, W1, ... , W15, where W0 is the left-most word.*
*b. For t = 16 to 79 let Wt = S1(Wt-3 XOR Wt-8 XOR Wt- 14 XOR Wt-16).*
*c. Let A = H0, B = H1, C = H2, D = H3, E = H4.*
*d. For t = 0 to 79 do*
TEMP = S5(A) + ft(B,C,D) + E + Wt + Kt;
E = D;
D = C;
C = S30(B);
B = A;
A = TEMP;
*e. Let H0 = H0 + A, H1 = H1 + B, H2 = H2+ C, H3 = H3 + D, H4 = H4 + E.*

After processing Mn, the message digest is the 160-bit string represented by the 5 words H0 H1 H2 H3 and H4.

### 3. PROPOSED SHA-192 ALGORITHM

The proposed SHA-192 is another improved version in SHA family. It may be used to hash message, M having a length of l bits, where $0<l<2^{64}$. The algorithm uses, Six working variables of 32 bits each, A hash value of six 32-bit words. The final result of SHA-192 is the 192 bit message digest. The words of the message schedule are labeled W0, W1, W2…W79. The six working variables are labeled A,B,C,D,E and F. The words of the hash value are labeled H0(i),...,which will hold the initial hash value, and is replaced by each successive intermediate hash value(after each message block is processed)and ending with final hash value H(N).

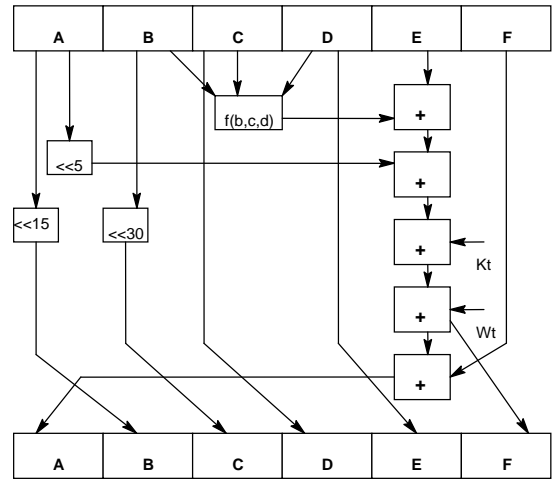

**Figure 4. SHA-192 compression function**

*3.1. SHA-192 preprocessing*

The padding and appending of bits are done same as for MD5 and SHA-1. Before processing any blocks, the {Hi} are initialized as follows (in hexadecimal):

H0 = 67452301, H1 = EFCDAB89, H2 = 98BADCFE
H3 = 10325476, H4 = C3D2E1F0, H5 = F9B2D834.

The compression function of SHA-192 has been illustrated in Fig.4.

*3.2. SHA-192 hash computation*

A sequence of logical functions f0, f1,..., f79 is used in the SHA-192. Each ft, $0 <= t <= 79$, operates on three 32-bit words B, C, D and produces a 32-bit word as output. ft(B,C,D) is defined as follows, for words B, C, D,

*ft(B,C,D) = (B and C) or ((not B) and D), for $0 <= t <= 19$*
*ft(B,C,D) = B xor C xor D, for $20 <= t <= 39$*
*ft(B,C,D) = (B and C) or (B and D) or (C and D), for $40 <= t <= 59$*
*ft(B,C,D) = B xor C xor D, for $60 <= t <= 79$*

A sequence of constant words K(0), K(1), ... , K(79) is used in the SHA-1. In hex these are given by

Kt = 5A827999 ( $0 <= t <= 19$)
Kt = 6ED9EBA1 ($20 <= t <= 39$)
Kt = 8F1BBCDC ($40 <= t <= 59$)
Kt = CA62C1D6 ($60 <= t <= 79$)

Now M1, M2... Mn is processed. To process Mi, we proceed as follows:

*a. Divide Mi into 16 words W0, W1... W15, where W0 is left-most word.*

*b. For t= 0 to 15 Wt = Mi*
   *For t = 16 to 79 let Wt = S1(Wt-3 XOR Wt-8 XOR Wt- 14 XOR Wt-16).*

*c. Let A = H0, B = H1, C = H2, D = H3, E = H4, F = H5.*

*d. For t = 0 to 79 do*
   TEMP1 = S5(A) + ft(B,C,D) + E + Wt + Kt;
   TEMP2 = S5(A) +A + ft(B,C,D) + E + Wt + Kt+F;
   E = D;
   D = C;
   C = S30(B);
   B = S15(A);
   F = TEMP1;
   A= TEMP2
*e. Let H0 = H0 + A, H1= H1 + B, H2 = H2+ C, H3 = H3 + D,*
   H4 = H4 + E, H5 = H5 + F.

After processing Mn, the message digest is the 160-bit string represented by the 6 words H0 H1 H2 H3 H4 and H5.

4. UNIFIED ARCHITECTURE OF MD-5 AND SHA-192

Many architecture has been used to implement these hash function individually in hardware [8]-[15]. The proposed architecture figure 5. has two built in hash function say MD5 and the proposed SHA-192. Both the algorithms in same architecture so that it can operate for one function one time and for other function next time. In the case of the MD5 operation the data transformation four inputs/outputs B,C,D,E of each one of the four Data Transformation Rounds. The input/output named A,F is not used , for this hash function operation (MD5). This is due to the fact that MD5 processes on 128-bit blocks (4x32-bit) transformation blocks, instead of the 192-bit blocks that are used in SHA-192. The four Data Transformation Rounds are similar, but its one performs a different operation. MA components indicate modulo addition 232, while the shifters components define left shift rotations of the input data block[8]. The Data Transformation Round I operation is based on a Nonlinear Function i transformation of the three of BIn, CIn, and DIn, inputs. Then, this result is added to the fourth input EIn with the input data block and the constant. That result is rotated to the right and the rotated output data are added with the input DIn. The each Data Transformation Round, which perform the digital logic transformation according to eq.uations.The Hash the Function Core shown in fig 5 can be used alternatively for the operation SHA-192 hash function also. The data transformation unit and the data transformation rounds process the data in a different way, compared with MD5 operation mode, in order the Hash Function Core to perform efficiently as SHA-192.For the SHA-192 operation each Data Transformation Round operates on all the six 32-bit variables (inputs/outputs) and this is one of the basic differences compared with MD5 mode. Thus the combined architecture of MD-5 and SHA-192 results in reduced hardware utilization compared to the individual implementation of MD-5 and SHA-1.

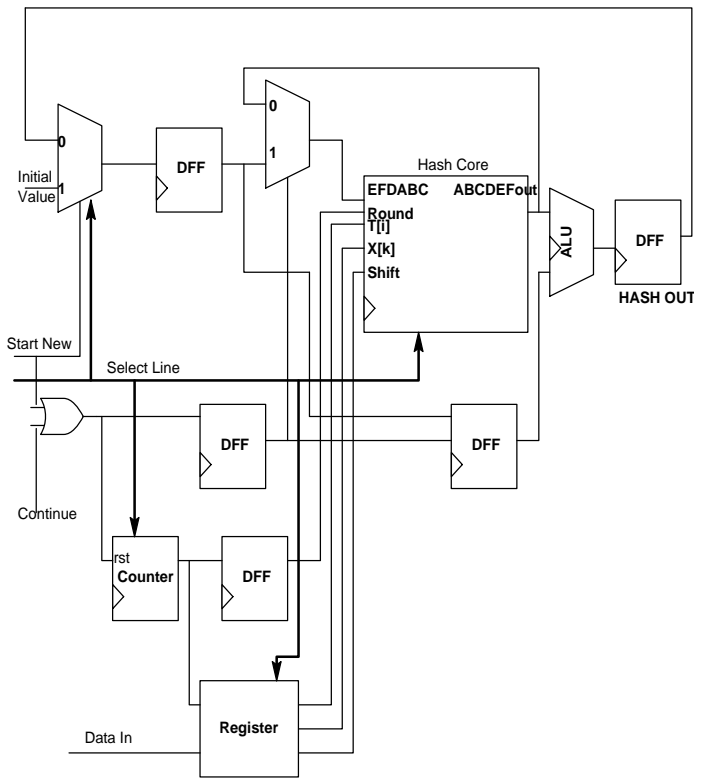

**Figure 5 MD5 and SHA-192 Unified Achitecture**

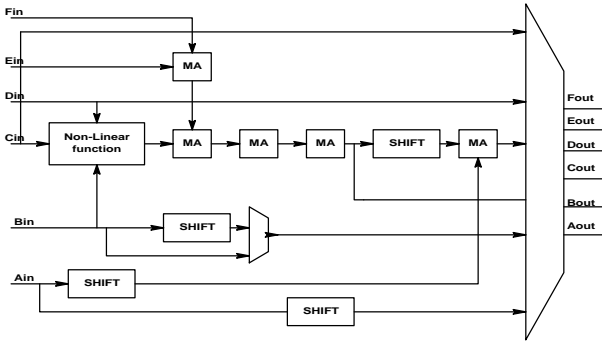

**Figure 6. Data Transformation for Combined Architecture**

The data transformation for combined hash computation is described in Fig 6. In the hardware realization, a select line is inserted which selects the functionality of appropriate algorithm at each block.

The data transformation and hash output of both individual and combined hash computations are shown in Fig 7 and 8. In the hardware realization, a select line is inserted which selects the functionality of appropriate algorithm at each block.

## 5. Results and discussion

The hardware architecture is implemented in verilog, and synthesis is performed with xilinx ise 9.2i.virtex ii kit is used for downloading the synthesized code. The power analysis is done using Synopsys-Design vision. The synthesis result for individual implementation of MD5 and SHA-1 is tabulated in table1 and 2. For the implementation, FPGA device 2v4000bf957-6 was used and the achieved operating frequency is equal to 57.36 MHZ and the system allocated area are 162 I/Os, 724 Function generators and 406 CLBs and 298 Dffs are utilized. In table individual implementation of SHA-1 is summarized. The achieved operating frequency is equal to 83.801 MHZ and the system allocated area are 194 I/Os, 2349 Function generator and 1333 CLBs and 1257 Dffs are utilized.

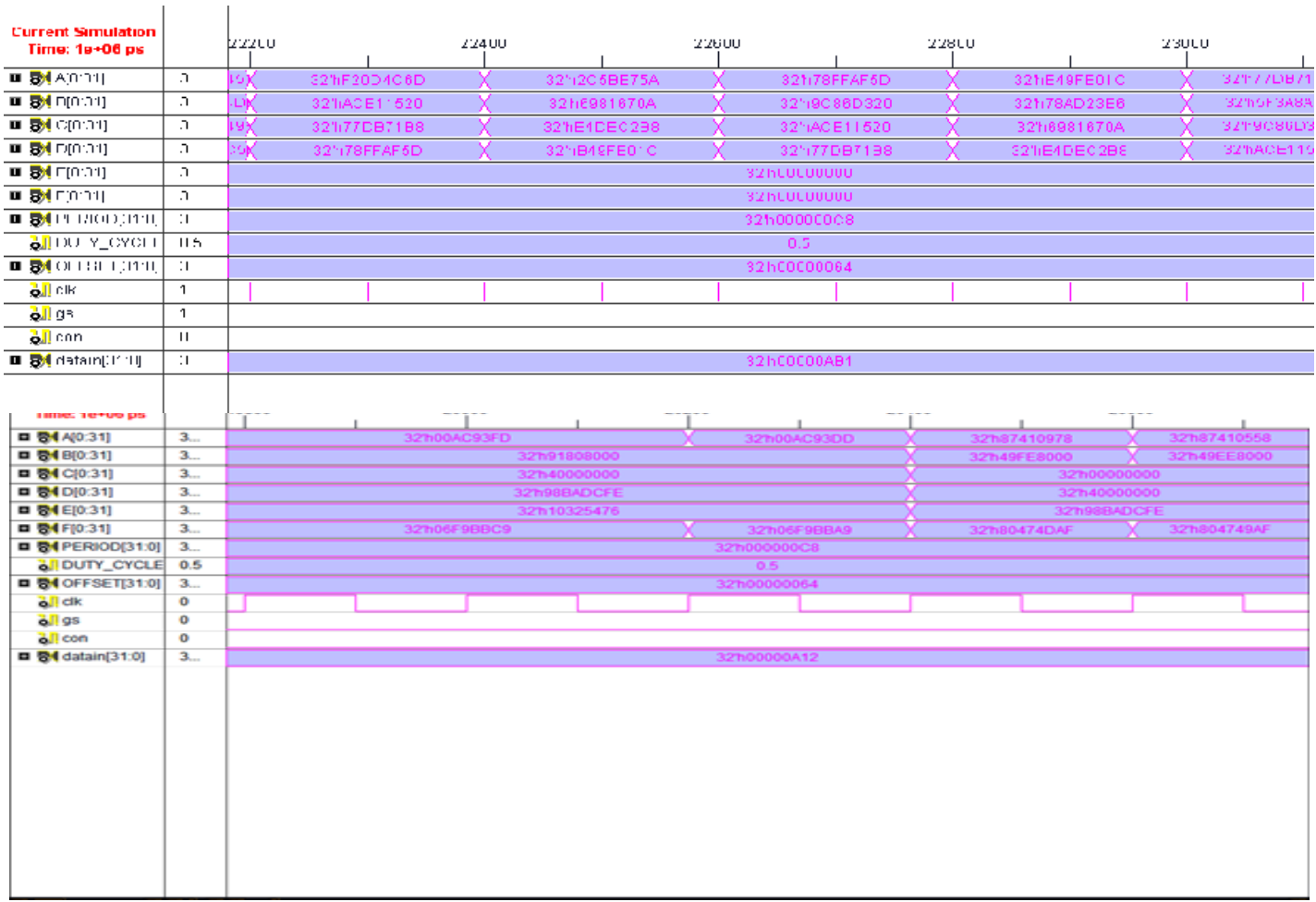

**Figure 7 Hash output of MD5 and SHA-192 in Individual and combined archite cture**

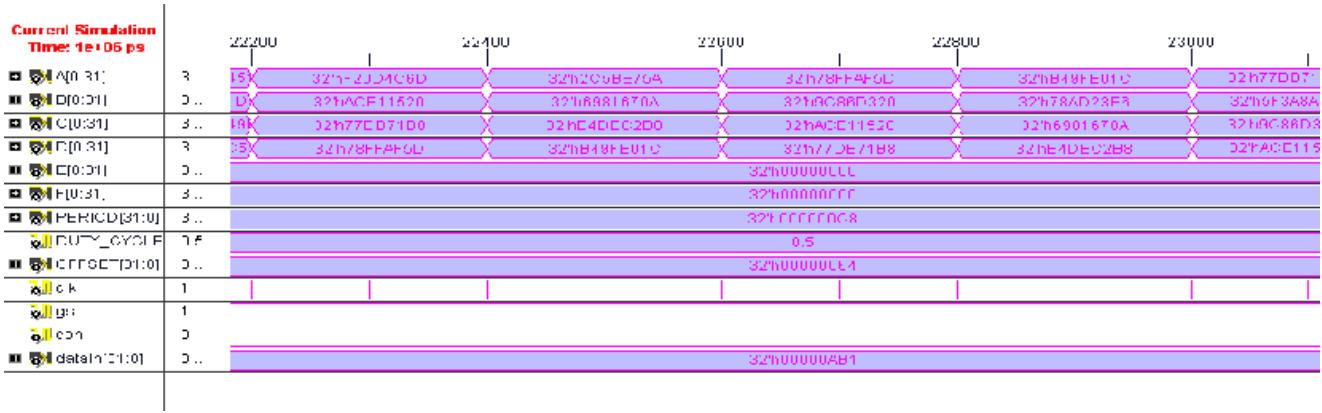

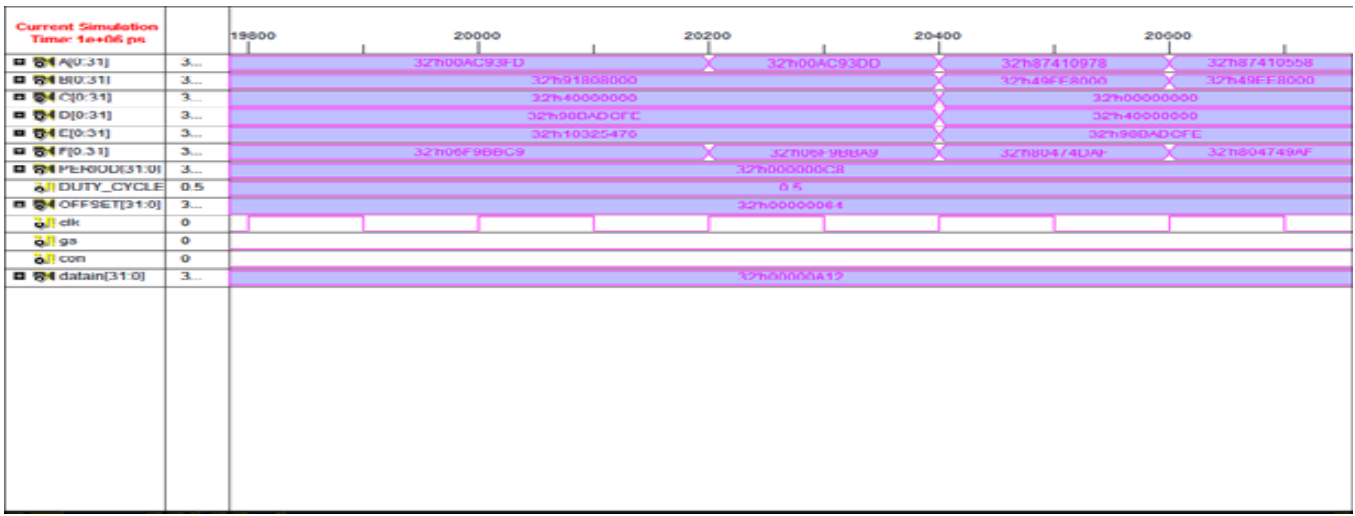

**Figure 8 Hash output of MD5 and SHA-192 in combined archite cture**

TABLE 1. HARDWARE UTILIZATION OF MD5

| FPGA device : 2v4000bf957-6 | | |
|---|---|---|
| *Allocated area* | *Used/Available* | *Utilization* |
| *I/Os* | 162/684 | 23% |
| *Fun. Generators* | 724/46080 | 1% |
| *CLB Slices* | 406/23040 | 1% |
| *Dffs and Latches* | 298/46080 | 0% |
| *frequency* | 57.36MHZ | |
| *Power consumption* | 4.55MW | |

TABLE 2. HARDWARE UTILIZATION OF SHA192

| FPGA device : 2v4000bf957-6 | | |
|---|---|---|
| *Allocated area* | *Used/Available* | *Utilization* |
| *I/Os* | 194/684 | 28% |
| *Fun.Generators* | 2349/46080 | 5% |
| *CLB Slices* | 1333/23040 | 5% |
| *Dffs and Latches* | 1257/46080 | 2% |
| *Frequency* | 83.801 MHZ | |
| *Power Consumption* | 15.49 mW | |

In table 3. Hardware utilization of unified architecture is summarized and message digests of MD5 and SHA-192 shown in fig 7 and 8. The comparative study shows that unified architecture utilize less area than individual structure and also consumes less power.

TABLE 3. HARDWARE UTILIZATION OF UNIFIED ARCHITECTURE

| FPGA device : 2v4000bf957-6 | | |
|---|---|---|
| *Allocated area* | *Used/Available* | *Utilization* |
| *I/Os* | 195/684 | 28% |
| *Fun.Generators* | 1275/46080 | 2% |
| *CLB Slices* | 757 / 23040 | 3% |
| *Dffs and Latches* | 1033/46080 | 2% |
| *frequency* | 105.67 MHZ | |
| *Power Consumption* | 7.092mW | |

From the tabulation, it could be inferred that the device utilization is less in unified architecture compared with the individual implementation of MD-5 and SHA-1. The unified architecture of MD-5 and SHA-192 proved to consume less power and also efficient in computing the hash.

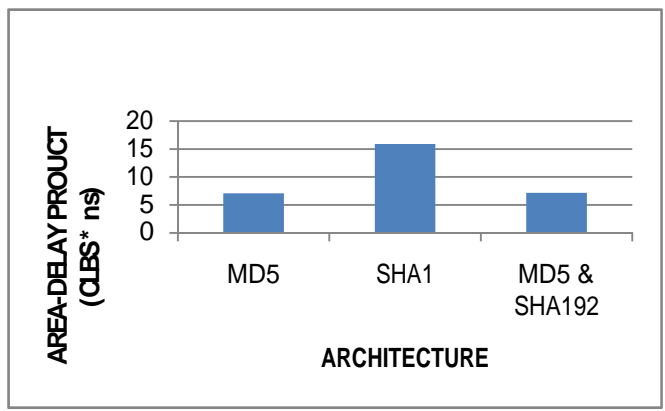

Figure 9. Area Delay Product Comparison.

Fig.9 shows the area delay product comparison of individual architectures and the unified architecture. From which it could be inferred that the area utilization of proposed combined architecture is less than available logic of FPGA chip used. The requirement of portability of mobile phones and hand held devices places severe restriction on power consumption. In proposed architecture low power design techniques is used to reduce the power consumption.

## CONCLUSION

In this work, VLSI architecture of the integrity unit for the reconfigurable receiver is presented. The propsed architecture is reconfigurable in the sense that operates either to give MD-5 hash or the SHA-192 message digest. It guarantees high security level in reconfigurable receivers requiring data integrity and message authentictaion. The comparisons of synthesis results, proved that the proposed integrity unit is better, compared with the individual implementation of the hash algorithms. The power consumption is also proved to be applicable for the reconfigurable receiver terminals.the introduced Integrity unitcan be sed in all types of SHA-1 ,MD5 application with high speeed demands and with high level scurity needs at the same time.

L.Thulasimani has obtained her BE and ME degree from Coimbatore Institute of Technology, India in 1998 and 2001 respectively. She has started her teaching profession in the year 2001 in PSNA engineering college, Dindigul. At present she is an Lecturer in department of Electronic and Communication Engineering in PSG college of Technology, Coimbatore .She has published 4 research papers in International and National conferences. She is a part time Ph.D research scalar in Anna University Chennai. His areas of interest are Wireless security, Networking and signal processing. She is a life member of ISTE.

Dr. M. Madheswaran has obtained his Ph.D. degree in Electronics Engineering from Institute of Technology, Banaras Hindu University, Varanasi in 1999 and M.E degree in Microwave Engineering from Birla Institute of Technology, Ranchi, India. He has started his teaching profession in the year 1991 to serve his parent Institution Mohd. Sathak Engineering College, Kilakarai where he obtained his Bachelor Degree in ECE. He has served KSR college of Technology from 1999 to 2001 and PSNA College of Engineering and Technology, Dindigul from 2001 to 2006. He has been awarded Young Scientist Fellowship by the Tamil Nadu State Council for Science and Technology and Senior Research Fellowship by Council for Scientific and Industrial Research, New Delhi in the year 1994 and 1996 respectively. His research project entitled "Analysis and simulation of OEIC receivers for tera optical networks" has been funded by the SERC Division, Department of Science and Technology, Ministry of Information Technology under the Fast track proposal for Young Scientist in 2004. He has published 120 research papers in International and National Journals as well as conferences. He has been the IEEE student branch counselor at Mohamed Sathak Engineering College, Kilakarai during 1993-1998 and PSNA College of Engineering and Technology, Dindigul during 2003-2006. He has been awarded Best Citizen of India award in the year 2005 and his name is included in the Marquis Who's Who in Science and Engineering, 2006-2007 which distinguishes him as one of the leading professionals in the world. His field of interest includes semiconductor devices, microwave electronics, optoelectronics and signal processing. He is a member of IEEE, SPIE, IETE, ISTE, VLSI Society of India and Institution of Engineers (India).